\documentclass[cameraready]{Interspeech} 

\title{Tracing the Origins: Legacy Codec Identification in Neural Audio Transcoding} 

\author[affiliation={1}, orcid=0000-0003-1753-8977, equalcontribution]{Wonje}{Heo}
\author[affiliation={2}, orcid=0009-0004-7868-4673, equalcontribution]{Shinee}{Youn}
\author[affiliation={1}, orcid=0009-0002-1687-7432]{Yooshin}{Kim}
\author[affiliation={1}, orcid=0009-0001-6122-9752]{Chuck}{Chae}
\author[affiliation={1}, orcid=0009-0004-7088-3846, correspondingauthor]{Donghoon}{Shin}

\address{
    $^1$ Department of Electrical Engineering and Computer Science (EECS), DGIST, Republic of Korea \\
    $^2$ School of Undergraduate Studies, DGIST, Republic of Korea
}

\email{\{heo\_wonje, Y.sini, yooshin0303, cocjr0208, dshin\}@dgist.ac.kr}

\keywords{audio forensic, neural audio codec, transcoding, multiple compression, source identification}

\usepackage{comment}
\usepackage{cite} 

\begin{document}

\maketitle

\begin{abstract}
Residual Vector Quantization (RVQ)-based neural audio codecs (NACs) enable high-fidelity audio distribution at unprecedentedly low bitrates through discrete token-based representations. However, this shift disrupts traditional forensics, as non-linear neural transcoding obscures the underlying traces of legacy compression. This study defines the forensic gap and proposes a Transformer-based framework designed to leverage the hierarchical and temporal dependencies inherent in RVQ sequences. By modeling inter-layer causal relationships and dynamic forensic significance, our model effectively disentangles superimposed artifacts from legacy-to-neural transcoding. Experimental results achieve 97\%+ accuracy for codec identification and robust joint identification performance across 32--128 kbps. These results demonstrate that traditional codec traces persist even after neural transcoding, supporting the feasibility and necessity of neural-codec-aware audio forensics.
\end{abstract}

\section{Introduction}

Digital audio has become a central multimedia medium in modern society, widely distributed through diverse platforms ranging from music streaming, podcasts, and broadcasting to social media. In this environment, compression technology is essential for efficiently storing and transmitting high-capacity raw audio signals. Consequently, audio codecs have evolved as key technologies that maximize data efficiency by combining psychoacoustic models, signal processing techniques, and information-theoretic compression \cite{mp3}.

Generally, traditional audio codecs process signals by removing perceptual redundancies based on human psychoacoustic models. During this process, specific structural and statistical artifacts—such as spectral cut-offs, quantization noise, and band-wise distortions—are inherently left behind depending on the design algorithm of each codec \cite{mp3, TradCodecArtifacts}. Audio forensics is a field that analyzes these subtle artifacts to identify processing history or infer potential tampering. Since audio manipulation or redistribution inevitably involves decoding and re-compression, codec-specific artifacts accumulate in the signal \cite{Bitrate1_Transformer, Bitrate2_FakeQuality}. By analyzing these accumulated traces, forensic analysts can identify not only the authenticity but also the source codec type and potential distribution \cite{AudioForensicReview}. Accordingly, research to identify single and multiple compression histories has been actively pursued \cite{TradCodecGMM, TradCodecHicsonmez}.

However, recent rapid advancements in deep learning have accelerated the emergence of Residual Vector Quantization (RVQ)-based Neural Audio Codecs (NACs), such as SoundStream, EnCodec, and DAC \cite{LyraV2, EnCodec, DAC}. By converting continuous audio signals into discrete token units, these models signify a fundamental change in the digital audio distribution pipeline, going beyond simple improvements in compression efficiency. This shift, however, creates a new security gap which challenges traditional forensic techniques based on legacy linear signal processing \cite{EnvId_failed_EnCodec, ASV_failed_EnCodec}. In particular, research that views NACs as a transmission medium and validates the effectiveness of traditional codec-based forensics—or proposes specialized analysis models for this new channel—has not been sufficiently addressed.

To bridge this forensic gap, we propose a novel analysis framework specifically designed to recover legacy compression traces directly from the discrete neural token domain. Rather than treating NACs as an impenetrable channel, our approach exploits the hierarchical nature of RVQ to disentangle superimposed artifacts. The primary contributions of this work are as follows: First, we address and systematically investigate the forensic problem of legacy-to-neural transcoding; Second, we introduce a Transformer-based architecture that captures hierarchical inter-layer dependencies and temporal signatures within RVQ sequences; Third, we demonstrate through extensive experiments that legacy codec artifacts remain inherently detectable even after neural re-compression. This study establishes a robust foundation for verifying audio integrity and tracking content provenance in the era of neural-driven distribution.
\section{Related Work}
\label{sec:related_work}

\subsection{Traditional Audio Forensics and Its Limitations in NACs}
Traditional audio forensics has extensively utilized statistical artifacts inherent in decoded waveforms or bitstreams to trace compression history \cite{mp3, TradCodecArtifacts}. Representatively, a Gaussian Mixture Model (GMM) was introduced to identify the source codec of singly compressed audio without prior knowledge of its internal structure \cite{TradCodecGMM}. Similarly, a Support Vector Machine (SVM) was employed to distinguish between single and double compressions based on encoded byte-stream information \cite{TradCodecHicsonmez}. However, these established methods predominantly rely on linear signal processing assumptions applied to continuous waveforms or deterministic bitstreams, which are significantly compromised by the non-linear transformation into discrete tokens within NAC environments. Empirical studies in environment identification \cite{EnvId_failed_EnCodec} and anti-spoofing \cite{ASV_failed_EnCodec} have consistently reported a performance collapse of traditional forensic techniques when applied to audio transcoded by NACs.

\begin{figure*}[t]
    \centering
    \includegraphics[width=\textwidth]{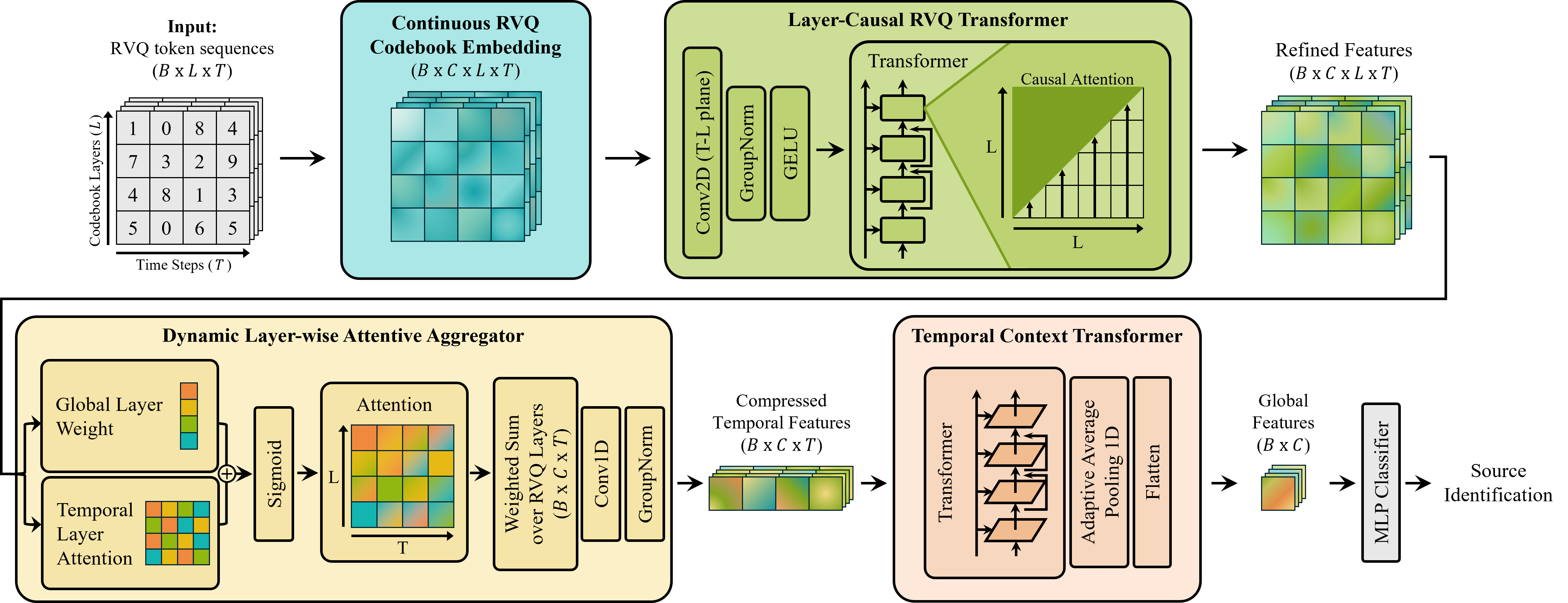}
    \caption{Overall architecture of the proposed model.}
    \label{fig:architecture}
\end{figure*}

\subsection{Forensic Gap in Neural Audio Codecs}
The advent of RVQ-based NACs, led by SoundStream and followed by EnCodec and DAC \cite{LyraV2, EnCodec, DAC}, signifies a paradigm shift in audio representation. By converting audio into discrete tokens, these models introduce complex, non-linear artifacts that differ fundamentally from legacy ones. Consequently, recent academic efforts have focused on exploring characteristics unique to NACs, such as comparing artifacts between models or verifying idempotence \cite{NeuralCodecAnalysis, NeuralCodecIdempotence}. However, in scenarios where a legacy codec signal is transcoded by a NAC, the new artifacts generated by RVQ tokens are superimposed onto the traces of the legacy codec, and the feasibility of source identification under this composite modulation remains unverified. While NACs are drawing attention as a next-generation distribution pipeline due to their high fidelity at low bandwidths \cite{RVQ_IsEfficient, RVQ_Enhanced}, academic interest remains predominantly focused on their utility in downstream generative tasks like Text-to-Speech (TTS) \cite{RVQ_TTS}. Consequently, a forensic framework that treats NACs as a transmission channel to resolve this artifact superposition and trace the original signatures of legacy codecs remains unaddressed.

\section{Proposed Method}
\label{sec:proposed_method}

We propose a forensic framework to identify the lineage of legacy source codecs within the discrete neural token domain. As illustrated in Figure~\ref{fig:architecture}, the framework processes RVQ token sequences $\mathcal{X} \in \mathbb{R}^{B \times L \times T}$, where $B, L, \text{and } T$ denote the batch size, the number of codebook layers, and the sequence length (time steps), respectively. These discrete indices are first mapped into a continuous embedding space of dimension $C$ using pre-trained NAC codebooks, yielding embedded representations $\mathbf{E} \in \mathbb{R}^{B \times C \times L \times T}$. To disentangle the superimposed artifacts from legacy-to-neural transcoding, the framework incorporates three specialized modules targeting the hierarchical, selective, and temporal characteristics of RVQ representations.

First, the \textbf{Layer-Causal RVQ Transformer (LCR-Trans)} models inter-layer dependencies through causal-masked attention. Second, the \textbf{Dynamic Layer-wise Attentive Aggregator (DLAA)} dynamically estimates the contribution of each codebook layer, aggregating them into a unified temporal feature. Finally, the \textbf{Temporal Context Transformer (TC-Trans)} captures codec-specific temporal signatures, such as pre-echo handling and bandwidth transitions. The resulting global feature is then fed into an MLP classifier for source identification.

\subsection{Layer-Causal RVQ Transformer (LCR-Trans)}
LCR-Trans is designed to leverage the hierarchical nature of RVQ, where the quantized output of a preceding codebook layer directly influences the representation of subsequent layers. This architecture is predicated on the observation that legacy codec artifacts are not merely present within isolated tokens but are inherently encoded as dependencies across these hierarchical layers. To capture these signatures, the module first employs a 2D convolutional layer to extract local correlations across neighboring time steps and codebook layers on the $T$–$L$ plane. The resulting features are refined via Group Normalization and GELU activation. Finally, a Transformer encoder models codec-specific inter-layer dependencies using causal-masked attention along the codebook-layer axis. This design effectively captures the hierarchical structure while safeguarding against information leakage from future layers, ensuring the integrity of the causal relationship.

\subsection{Dynamic Layer-wise Attentive Aggregator (DLAA)}
The DLAA module exploits the non-uniform forensic significance of codebook layers, as certain quantization levels inherently carry more distinctive traces of the legacy source codec depending on its bit-allocation or spectral shaping strategies. To reflect this, DLAA dynamically estimates the relative importance of each layer by deriving two distinct weighting factors from the LCR-Trans output: a Global Layer Weight, which encodes the overall layer-wise importance across the entire temporal span, and a Temporal Layer Attention, which captures time-varying layer significance at each time step $t$. These terms are combined via a sigmoid function to produce a dynamic attention map $\mathcal{A} \in \mathbb{R}^{B \times 1 \times L \times T}$. Using this map, the input features are reweighted and aggregated via a weighted sum along the layer axis, followed by a 1D convolution for channel mixing and feature refinement. This process yields compressed temporal features $\mathbf{Z} \in \mathbb{R}^{B \times C \times T}$ that prioritize the most forensically informative layers.

\subsection{Temporal Context Transformer (TC-Trans)}
Legacy codecs exhibit fundamental differences in handling time-varying acoustic events, such as pre-echo suppression and transient processing. These variations are encoded as unique contextual signatures along the temporal axis of the RVQ sequence. TC-Trans captures these patterns using a Transformer encoder that models long-range temporal dependencies across the aggregated feature sequence. The resulting features are then summarized into a global feature vector $\mathbf{v} \in \mathbb{R}^{B \times C}$ through Adaptive Average Pooling. This vector, which integrates both the hierarchical inter-layer relationships and temporal signatures, provides a robust and comprehensive representation for the final MLP-based source identification.

\section{Experimental Evaluation}
\label{sec:experimental_evaluation}

\subsection{Dataset}
\label{ssec:dataset}
We constructed a large-scale dataset based on the VCTK corpus \cite{VCTK}. Speech samples were first compressed using legacy codecs via FFmpeg and subsequently processed by a pretrained 48 kHz EnCodec model\cite{EnCodec}. To establish a reference condition, clean samples were processed directly by EnCodec without prior legacy compression. Since the 48 kHz EnCodec model requires stereo input, mono signals were converted to dual-mono format.

To simulate diverse real-world scenarios, we selected five ubiquitous codecs (\textbf{MP3 \cite{ref_MP3}, AAC \cite{ref_AAC}, Opus \cite{ref_OPUS}, Vorbis \cite{ref_VORBIS}, and G.711 $\mu$-law \cite{ref_G711}}), covering the majority of modern streaming traffic and ensuring algorithmic diversity. Specifically, MP3, AAC, and Vorbis employ MDCT-based transform coding, Opus uses hybrid linear prediction, and G.711 performs waveform coding. Signals were encoded at four bitrates (\textbf{32, 64, 96, and 128 kbps}) prior to neural transcoding, enabling evaluation across varying degradation levels. For Vorbis, we used the corresponding VBR quality levels q-2, q0, q2, and q4. We adopted a speaker-disjoint split to prevent speaker leakage, dividing speakers into 80\% training, 10\% validation, and 10\% test sets.

\begin{table}[h]
  \centering
  \caption{Target codecs and bitrate configurations}
  \label{tab:codecs}
  \resizebox{\columnwidth}{!}{%
  \begin{tabular}{l|c|l}
    \hline
    \textbf{Codec} & \textbf{Bitrates / Quality} & \textbf{Algorithm Type} \\
    \hline
    MP3 \cite{ref_MP3}    & 32, 64, 96, 128 kbps & MDCT / Psychoacoustic \\
    AAC \cite{ref_AAC}    & 32, 64, 96, 128 kbps & MDCT / TNS \\
    Opus \cite{ref_OPUS}   & 32, 64, 96, 128 kbps & Hybrid (SILK + CELT) \\
    Vorbis \cite{ref_VORBIS} & q-2, q0, q2, q4 ($\approx$ 32--128 kbps) & MDCT / VBR \\
    G.711 \cite{ref_G711}  & 64 kbps (Fixed)  & Logarithmic PCM \\
    \hline
  \end{tabular}
  }
\end{table}

\subsection{Experimental Setup}
The model is trained for 25 epochs using the AdamW optimizer with a batch size of 16. We set the initial learning rate to $5 \times 10^{-6}$ and weight decay to $1 \times 10^{-4}$. A linear warmup is applied for the first 5 epochs, followed by a ReduceLROnPlateau scheduler for learning rate adjustment. The optimal model is selected based on the highest validation Macro-$F_1$ score. For evaluation, we report both overall accuracy and Macro-$F_1$ on the speaker-disjoint test set.

\subsection{Codec Identification under Fixed Bitrate}
\begin{table}[t]
  \centering
  \caption{Codec identification performance under fixed bitrate}
  \label{tab:codec_identification}
  \begin{tabular}{lcccc}
    \hline
    \textbf{Metric} & \textbf{32 kbps} & \textbf{64 kbps} & \textbf{96 kbps} & \textbf{128 kbps} \\
    \hline
    Accuracy  & 99.99 & 99.70 & 98.36 & 97.32 \\
    Macro-F1  & 99.99 & 99.70 & 98.36 & 97.32 \\
    \hline
  \end{tabular}
\end{table}

Table~\ref{tab:codec_identification} presents the codec identification results under fixed bitrate conditions. The task configuration varied by bitrate: for 64 kbps, a 6-class identification was performed, encompassing five legacy codecs and clean reference. For the 32, 96, and 128 kbps settings, a 5-class identification was conducted, excluding G.711 as it operates exclusively at 64 kbps.

Overall, the model demonstrates robust performance across all bitrate configurations, consistently exceeding 97\% in both accuracy and $F_1$ score. Notably, at 32 kbps and 64 kbps, the model achieves 99.99\% and 99.70\%, respectively, representing near-perfect identification. A clear trend emerges where identification performance improves as the bitrate decreases. This is explained by the observation that more aggressive compression at lower bitrates—particularly at 32 kbps—introduces more pronounced codec-specific traces into the signal. These artifacts, ranging from severe spectral masking to unique quantization noise patterns, provide highly discriminative cues that are effectively captured by the proposed hierarchical and temporal modules.

\subsection{Bitrate Classification under Fixed Codec}
\label{sec:bitrate_classification}
\begin{table}[t]
  \centering
  \caption{Bitrate classification performance under fixed codec}
  \label{tab:bitrate_classification}
  \begin{tabular}{lcccc}
    \hline
    \textbf{Metric} & \textbf{MP3} & \textbf{AAC} & \textbf{Opus} & \textbf{Vorbis} \\
    \hline
    Accuracy  & 84.43 & 99.97 & 71.01 & 99.65 \\
    Macro-F1  & 84.40 & 99.97 & 70.60 & 99.65 \\
    \hline
  \end{tabular}
\end{table}

Table~\ref{tab:bitrate_classification} summarizes the bitrate classification performance across four bitrates (32, 64, 96, and 128 kbps). Both AAC and Vorbis exhibit exceptional performance, achieving accuracies exceeding 99\%. In contrast, MP3 and Opus show comparatively lower performance.

A detailed analysis of the confusion matrix reveals that MP3 and Opus encounter significant difficulties in distinguishing between 96 kbps and 128 kbps. This is likely due to artifact convergence as the bitrate approaches transparency. In contrast, the 32 kbps introduces a pronounced degradation gradient that distinguishes it from higher bitrates. The aggressive spectral masking and coarse quantization noise inherent at 32 kbps act as highly discriminative signatures, which our model effectively captures to achieve near-perfect identification. For Opus, this discrimination is further aided by its adaptive switching mechanism; while 64 kbps and above primarily rely on the CELT layer, the 32 kbps configuration often invokes hybrid coding strategies, providing additional forensic cues that prevent confusion with its high-bitrate counterparts.

\subsection{Joint Codec and Bitrate Identification}
\begin{figure*}[!t]
    \centering
    \includegraphics[width=\textwidth]{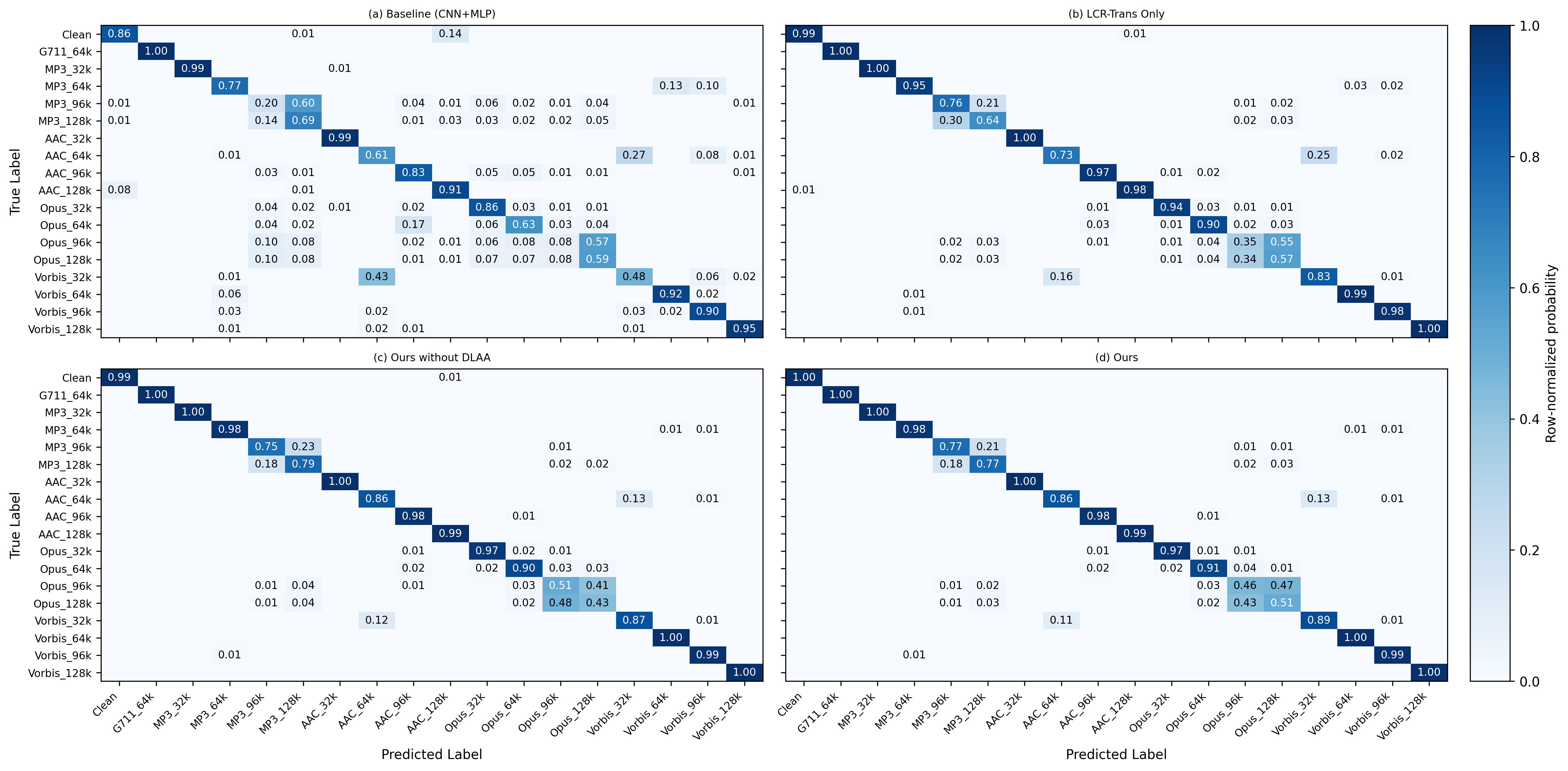}
    \caption{Row-normalized confusion matrices of the 18-class joint identification task across model variants. The proposed full model shows the strongest diagonal dominance, while confusion between MP3 and Opus at 96 and 128 kbps persists across configurations.}
    \label{fig:confusion_matrix}
\end{figure*}

\begin{table}[!t]
  \centering
  \caption{Performance comparison and ablation study on the 18-class joint identification task}
  \label{tab:18class_ablation}
  \resizebox{\columnwidth}{!}{%
  \begin{tabular}{lcc}
    \hline
    \textbf{Model} & \textbf{Accuracy} & \textbf{Macro-F1} \\
    \hline
    Baseline (CNN+MLP) & 73.71 & 72.18 \\
    LCR-Trans Only & 86.60 & 86.46 \\
    DLAA Only & 81.82 & 81.63 \\
    TC-Trans Only & 84.85 & 84.72 \\
    Ours without LCR-Trans & 86.88 & 86.80 \\
    Ours without DLAA & 88.97 & 88.92 \\
    Ours without TC-Trans & 87.98 & 87.87 \\
    Ours & \textbf{89.34} & \textbf{89.31} \\
    \hline
  \end{tabular}
  }
\end{table}
Finally, we conducted an 18-class identification task to jointly determine both codec and bitrate across the entire dataset. To the best of our knowledge, this is the first study to address such a joint identification setting, making direct comparisons with prior work challenging. To establish a reference, we defined a minimal baseline comprising RVQ embedding, a lightweight convolutional projection, global pooling, and an MLP classifier. In addition, we evaluated two sets of controlled variants: three single-module models, each retaining only one of the proposed components (LCR-Trans, DLAA, or TC-Trans), and three ablation models obtained by removing one module at a time from the full architecture. In total, eight models---including the minimal baseline and the full proposed model---were evaluated.

Table~\ref{tab:18class_ablation} presents the results. The proposed architecture achieves the highest performance, outperforming the CNN-based baseline by over 15\%. Furthermore, performance exhibits a clear upward trend as each module is incrementally integrated, indicating that all three modules contribute meaningfully to the model’s overall effectiveness. Among the three modules, LCR-Trans contributes the most to performance improvements.

This trend is further corroborated by the confusion matrices shown in Figure~\ref{fig:confusion_matrix}. As the model approaches its full configuration, the diagonal dominance becomes more pronounced, reflecting corresponding improvements in identification accuracy. For MP3 and Opus, confusion persists between 96 kbps and 128 kbps, as also observed in Section~\ref{sec:bitrate_classification}; this also leads to a slight performance degradation for Opus at 64 kbps. A minor confusion is also observed between AAC at 64 kbps and Vorbis at 32 kbps, indicating that slight ambiguity may arise between distinct artifact profiles at lower bitrates. Apart from these cases, the model achieves over 97\% accuracy in most conditions, demonstrating robust joint identification capability.

\section{Conclusion}
\label{sec:Conclusion}
This work formally defines a forensic gap that emerges as NACs are integrated into modern digital audio distribution pipelines. Specifically, we formulate the problem of tracing the original legacy codec when audio, initially encoded with a traditional codec, undergoes transcoding through an RVQ-based NAC. This study represents the first systematic attempt to analyze codec identifiability in legacy-to-neural transcoding.

Using the proposed model, we conducted three types of experiments: fixed-bitrate codec identification, fixed-codec bitrate classification, and a comprehensive 18-class joint codec–bitrate identification. Under fixed bitrate conditions (32, 64, 96, and 128 kbps), the model achieved over 97\% accuracy across all cases. In bitrate classification for specific codecs, the model demonstrated outstanding performance for AAC and Vorbis, exceeding 99\% accuracy. However, Opus presented a greater challenge at higher bitrates due to its subtle compression artifacts, resulting in relatively lower performance. In the most demanding 18-class joint identification task, the proposed model achieved over 89\% accuracy, outperforming all baseline models. These results demonstrate that residual artifacts from legacy codecs remain detectable even after subsequent neural transcoding. A minor confusion between AAC at 64 kbps and Vorbis at 32 kbps further indicates that certain low-bitrate artifact profiles may partially overlap.

This work lays the foundation for forensic analysis in legacy-to-neural transcoding, and several directions remain open for future research. In particular, improving identification performance under high-bitrate conditions for codecs such as MP3 and Opus remains an important direction. Additionally, expanding the study to incorporate a broader range of legacy and NACs would further enhance the generalization capability and robustness of the proposed approach.

\section{Acknowledgements}
This work was partly supported by the National Research Foundation of Korea (NRF) grant funded by the Korea government(MSIT) (No.RS-2025-23324100) and Institute of Information \& communications Technology Planning \& Evaluation(IITP) grant funded by the Korea government (MSIT) (No.RS-2025-02305765).


\section{Generative AI Use Disclosure}
The authors used generative AI tools to assist with language polishing and minor wording refinement. All technical content, experimental design, analysis, and conclusions were developed and verified by the authors.
\bibliographystyle{IEEEtran}
\bibliography{mybib}

\end{document}